\RequirePackage{amsmath}
\documentclass[runningheads, a4, anonymous]{llncs}

\usepackage{graphicx}
\usepackage[utf8]{inputenc}
\usepackage[T1]{fontenc}
\usepackage[lined,boxed,noend]{algorithm2e}
\usepackage{amsmath}
\usepackage{url}
\usepackage{eurosym}
\usepackage{amsfonts}
\usepackage{float}
\usepackage{hyperref}
\usepackage[hyphenbreaks]{breakurl}

\usepackage{comment} 
\usepackage[moderate]{savetrees}

\newcommand{\myheader}[1]{{\vspace*{5pt}}\noindent{\bf {#1}}}

\begin{document}

\title{Boardroom Voting:\\ 
Verifiable Voting with Ballot Privacy\\
Using Low-Tech Cryptography in a Single Room}

\titlerunning{{\it Low-Tech Boardroom Voting}}

\author{Enka Blanchard\inst{1} 
\and Ted Selker\inst{2} 
\and Alan T. Sherman\inst{3}}


\institute{Digitrust, Loria, Universit\'{e} de Lorraine \email{Enka.Blanchard@gmail.com}, \url{www.koliaza.com}
\and University of California Berkeley and UMBC,\email{ted.selker@gmail.com}, \url{http://ted.selker.com/}
\and Cyber Defense Lab, University of Maryland, Baltimore County (UMBC),\email{sherman@umbc.edu}, \url{https://www.csee.umbc.edu/people/faculty/alan-t-sherman/}}



\maketitle              



\begin{abstract}


A {\it boardroom election} is an election that takes place in a 
single room --- the boardroom --- in which all voters can see and hear each other.
We present an initial exploration of boardroom elections 
with ballot privacy and voter verifiability 
that use only ``low-tech cryptography'' without using computers to mark or collect ballots.
Specifically, we define the problem, 
introduce several building blocks,
and propose a new protocol that combines these blocks in novel ways.
Our new building blocks include ``foldable ballots'' that can be rotated to
hide the alignment of ballot choices with voting marks,
and ``visual secrets'' that are easy to remember and use but hard to describe.
\newline \hspace*{0.25in} 
Although closely seated participants in a boardroom election have limited privacy, 
the protocol ensures that no one can  determine how others voted. 
Moreover, each voter can verify that their ballot was correctly 
cast, collected, and counted, without being able to prove how they voted, providing 
assurance against undue influence.
\newline \hspace*{0.25in} 
Low-tech cryptography is useful in situations where constituents do not trust 
computer technology, and it avoids the complex auditing requirements of
end-to-end cryptographic voting systems such as Pr\^{e}t-\`{a}-Voter.
This paper's building blocks and protocol are meant to be a proof of concept that might be tested for usability and improved.



\end{abstract}


\section{Introduction}
\label{sec:intro}

Much research on election technology has focused on mass elections conducted in person using precincts or kiosks, or 
at distance using mail-in ballots or the Internet. Edison's~\cite{Edison1869}
first attempt at a private voting solution is inappropriate as it was for congress where the public needs to know 
how each participant voted.\footnote{This work has been submitted to the IEEE for possible publication. Copyright may be transferred without notice, after which this version may no longer be accessible.}
Many important elections, however, require anonymity but take place where it is easy to see others work, such as around a conference table in a room.
For example, a board of directors might vote whether to adopt a new corporate policy;
a committee of professors might vote whether to grant tenure to a colleague; or shareholders might decide on a business action.
 
 
Such ``boardroom elections'' typically use paper ballots 
with limited to no guarantees of ballot privacy, outcome integrity, or coercion resistance. 
Sitting around a table, for example, it is hard to hide your vote from your neighbour.  
With sleight-of-hand, anyone handling the ballots might replace a ballot with a fraudulent one without detection.
There is no assurance that ballots were not modified prior to counting them. 
Although the boardroom setting presents challenges for ballot privacy, 
it also offers some advantages:  one could prevent non-voters from entering the room, and everyone in the room can observe each other.

For example, one of the authors recently participated in a tenure vote during which each of the 22 voters could easily see the ballot choices of nearby voters; anyone could see ballot marks through the folded paper ballots; and ballots were distributed and collected
in a chaotic fashion during which anyone could handle the blank and marked ballots.  
One left-handed voter marked their ballot with distinctive backward checks in red ink.
While simple paper ballot elections could be conducted more securely, usually they are not.

Scrutiny of boardroom election procedures goes back centuries, with a 1274 decree specifying the procedures for bishops to elect the next pope. But such procedures and modern proposals either lack ballot privacy or outcome integrity, or require advanced technology (e.g., complex cryptography carried out on computers). This paper focuses on low-tech solutions.

We present a protocol, BVP1, for such boardroom elections with ballot privacy and voter verifiability
that uses only ``low-tech cryptography'' without any computers.
Our simple low-tech paper-based solution simplifies the trust model and does not require the sophisticated cryptographic audits integral to most 
{\it End-to-End (E2E)} systems, such as Scantegrity~\cite{Carback2010ScantegrityIImunicipal,Carback2016ScantegrityVotingSystem,Chaum2008ScantegrityIIEnd,Chaum2009ScantegrityIIEnd} or  Pr\^{e}t-\`{a}-Voter~\cite{Khader2013ProvingPretVoter,Ryan2011PretVoterConfirmation,Ryan2009Pretvotervoter}. 
The independence from electronic tools also ensures limited cost and improved availability in a wide variety of settings, while assuaging widespread concerns about computer malfeasance. 

Well-known examples of low-tech cryptography include the following:
committing to a secret value by covering it with black photographic tape;
encrypting a message by locking it in a safe box; and
creating a unique unreproducible tamper-evident seal by randomly sprinkling many tiny sparkles in translucent epoxy glue~\cite{Simmons1992}. 


The protocol focuses on the following properties:
No one can determine how any individual voted, even when observing
 a voter marking their ballot from close proximity.
Each voter can verify that their ballot was correctly 
cast, collected, and counted.
No voter can prove to anyone else how they voted, providing 
assurance against undue influence.
Each voter can be convinced of any malfeasance involving their vote.
In the basic version of BVP1, the voter cannot prove such malfeasance to anyone else.  
We present a variation of BVP1 in which objecting voters can
prove such malfeasance at the cost of some degradation of ballot privacy.

Contributions include:
(1)~A definition of the boardroom voting problem.
(2)~New building blocks for boardroom elections, including
``foldable ballots'' that can be manipulated to obfuscate the alignment of ballot choices with voting marks, 
and ``visual secrets'' that are easy to remember but hard to describe.
(3)~A new  protocol for boardroom voting that offers ballot privacy and voter-verifiable outcome integrity.

We recognise that our proposals need to be tested for usability.  
We offer them, not as an ultimate solution, but with the
hope that this initial exploratory work will inspire others to seek additional solutions to boardroom voting.




\section{Low-Tech Boardroom Elections}
\label{sec:background}

A \textit{boardroom election} is an election that takes place with all voters present in a single room, which shall called the \textit{boardroom}. A crucial property of such elections is that all voters can see and hear each other. While there is no rigid maximum number of voters, a typical boardroom election might involve three  to forty voters. The election is administered by an untrusted voter or their untrusted assistants, also present in the room, which we shall call the {\it Election Authority (EA)}. The election begins and ends in the boardroom. The process might be supported by some materials, such as paper ballots, marking devices, tape, stamps, and other objects which can be  acquired in advance.

Solutions should be simple, afford ballot privacy, resist undue influence,
and provide outcome integrity verifiable by the voters present.  In particular, solutions should not require the use of complex technology, such as laptops or sophisticated cryptographic software.  These requirements do not exclude the use of cryptography, but require that any cryptography be carried out in a ``low-tech'' fashion (e.g., implementing a cryptographic commitment by covering a character string with black photographic tape).

The system should satisfy the security requirements of \textit{ballot privacy} and \textit{outcome integrity}. {Ballot privacy} means that, even with the cooperation of corrupt voters, no one should have the ability to link a marked ballot to the voter who cast it.  Ballot privacy protects against undue influence, including vote selling and coercion. {Outcome integrity}~\cite{Benaloh2015Endendverifiability} means that the voters can verify that (1)~they cast their ballot as intended; (2)~the ballots were collected as cast; and (3)~the ballots were counted as collected.  We distinguish between two types of outcome verifiability: \textit{Weak verifiability} means that a voter can convince themselves if outcome integrity is violated.  \textit{Strong verifiability} means that the voter can additionally convince others of such malfeasance. 

Ideally, the system should resist delay and disruption, and it should not be possible for a corrupt voter to convince other voters with a false claim of malfeasance (that is, the system should resist {\it discreditation attacks}).



\section{Assumptions and Adversarial Model}
\label{adversary}

The  assumptions and adversarial model, include
characteristics of the room and 
the adversary's motivations, capabilities, access, resources, and risk tolerance.


\subsection{Assumptions} 
\label{boardroom:assumptions}

Assume the boardroom has sufficient size, light, and acoustics that the voters can be all present in the room, see each other, and hear each other.  
Cameras and electronic devices --- including cell phones --- are not permitted, 
or at least their use is prohibited while the election is in progress.
Assume that no cameras are hidden or otherwise present in the room.  
Given that some of our building blocks provide some defence against cameras, 
this assumption can be revisited.
Similarly, assume that it is not possible to peer into the room from outside, for example, using a telescope aimed through a window.


The situation, however, is sufficiently cosy that each voter can see what nearby voters are doing or writing at their seat.  There can be a place in the room that offers privacy  ---  for example, by using a privacy screen  ---  where voters can go, one at a time, to carry out certain voting steps.
The only people present in the room are the {voters} and, possibly, a few people acting as the election authority.  



During the election, communications among people in the room are not allowed beyond those required for the election procedure.  It would be impossible, however, to prevent all such communications completely, possibly including ones sent through covert channels (e.g., hand gestures).  Assume that such illicit communications are either detected or have limited bandwidth. 

\subsection{Adversarial Model}

The adversary's goals may include any of the following:
influence the result of the election;
find out how certain voters voted;
prevent, delay, or discredit the election; or
frame a specific voter for trying to disrupt the election.

The adversary is covert and might be a voter or member of the election authority.
Multiple adversaries might act in concert, or each for a different  ---  and potentially conflicting  ---  goal. Regardless, the adversaries have complete knowledge of the election system and all procedures.  

To achieve their goals, the adversary has access to financial and technical resources.
Assume they have copies of the materials used in the election  ---  at least for materials that are not unique. They can try to bribe or coerce one or more of the voters.  
Because they are in the boardroom, they can also peer over other people's shoulders and look at what voters write and do.

To some limited extent, the adversary is capable of executing certain sleight-of-hand activities. For example, the adversary might drop two ballots into a ballot box instead of one without detection, or make a ballot vanish (e.g., into their sleeve). 
Such manoeuvres can affect the distribution or collection of physical materials, unless additional protections are enforced.

Assume that the adversary wishes not to be detected.  Thus, the adversary does not wish to reveal their malicious intentions, and a failed attack might lead to serious consequences (e.g., lost reputation, lingering doubts, loss of job, investigation).
Unlike electronic attacks, which might be carried out at a distance and be hard to trace, 
boardroom attacks by an adversary in the room might carry high risks.
Consequently, deterrence may play an important role.


\section{Previous Work}
\label{sec:previous}

Small-scale elections in a single room have been organised and studied for centuries, a prime example being the papal election. Its rules are still mostly based on the papal decree 
\textit{Ubi Periculum}~\cite{PGX1274UbiPericulum}, written in 1274, and made into canon law in 1298~\cite{Colomer1998Electingpopesapproval}. Although it describes in great detail the way the electors should interact with the outside world, and requires the winner to be elected by at least a two-thirds supermajority, 
it makes no mention of how the vote is to happen. 
More recent rulings forbid the presence of any audio-visual recording equipment~\cite{PJPII1996UniversiDominiciGregis}. 
They also establish some formal requirements, including ballot 
chain of custody and
ballot format (secret ballots, with explicit constraints on their size and design). These rules, however, do not address the issues of privacy and verifiability in the presence of a skilled adversary. 

There are some images and speculations about how ancient Greeks may have voted by dropping a pebble, a pottery bit, or a small bronze 
disk --- to which was attached a peg corresponding to the 
vote --- into a tall urn or urns, possibly creating an audible sound~\cite{Boegehold1963studyAthenianvoting,Canevaro2018MajorityRulevs}.
Although much remains unclear about how the ancient Greeks actually voted, we can imagine very attractive methods involving dropping pebbles into urns behind the protection of a privacy screen.
For example, some fraternal organizations vote on each new membership application by ``Blackballing''
in which, behind a privacy screen, each voter drops either one white ball or one black ball into
a collection box.

Today, boardroom voting commonly occurs in classrooms, faculty meetings, and
company management or shareholder meetings.
Intimidation and fraud are frequent~\cite{Barrett2009ElephantBoardroomCounting,Kahan2007hangingchadscorporate}. 
Within the past fifteen years, 
researchers have proposed several solutions, based on electronic means, including smartphones~\cite{Bergen2014mobileapplicationboardroom}, blockchains~\cite{McCorry2017SmartContractBoardroom}, 
authenticated communication channels~\cite{Groth2004EfficientMaximalPrivacy}, 
or insecure devices~\cite{Arnaud2013AnalysisElectronicBoardroom}. 
Such cryptographic solutions have attempted to improve efficiency~\cite{Kulyk2015EfficiencyEvaluationCryptographic} 
or add features such as 
decentralisation~\cite{McCorry2017SmartContractBoardroom}, robustness~\cite{Khader2012fairrobustvoting}, 
or the possibility of vote delegation~\cite{Kulyk2017EnablingVoteDelegation}.
Kahan and Rock~\cite{Kahan2007hangingchadscorporate} examined corporate voting in the United States from a legal perspective.
For these previous works, the main defining characteristic of boardroom voting is an election with a small number of voters.

Kiayias and Yung~\cite{Kiayias2002SelftallyingElections} explored self-tallying cryptographic voting methods that may be useful in the boardroom because they
offer strong ballot secrecy and simplified post-casting procedures.

Kulyk~\cite{Kulyk2015EfficiencyEvaluationCryptographic}
surveyed and compared cryptographic boardroom voting, assuming
a common network, the deployment of a public-key infrastructure,
and that each voter has an electronic device.
Kulyk  also compiled a list of useful cryptographic primitives and protocols and compared their computational complexity. 

Essex et al.~\cite{Aperio2008} present the integrity-verification
mechanism Aperio for use in minimally equipped secret paper-ballot elections.

Hao~\cite{Hao2018VerifiableClassroomVoting} studied ``classroom voting,'' 
where the most important requirements are
minimising the cost of election materials and 
using open-source software and 
readily available low-cost hardware.

Section~\ref{sec:comparison} also discusses a paper ballot system corresponding to what is often done in practice.




\section{New Building Blocks}
\label{sec:blocks}
\label{building_blocks}

We present three new building blocks for the physical protocol of Section~\ref{sec:protocol}:
foldable paper ballots, visual secrets, and parallel vote tallying.
For additional building blocks, see Section~\ref{sec:more-blocks}.


\subsection{Foldable Paper Ballots} \label{boardroom:foldedpaper}

\begin{figure}    
    \centering 
    \includegraphics[width=0.5\textwidth]{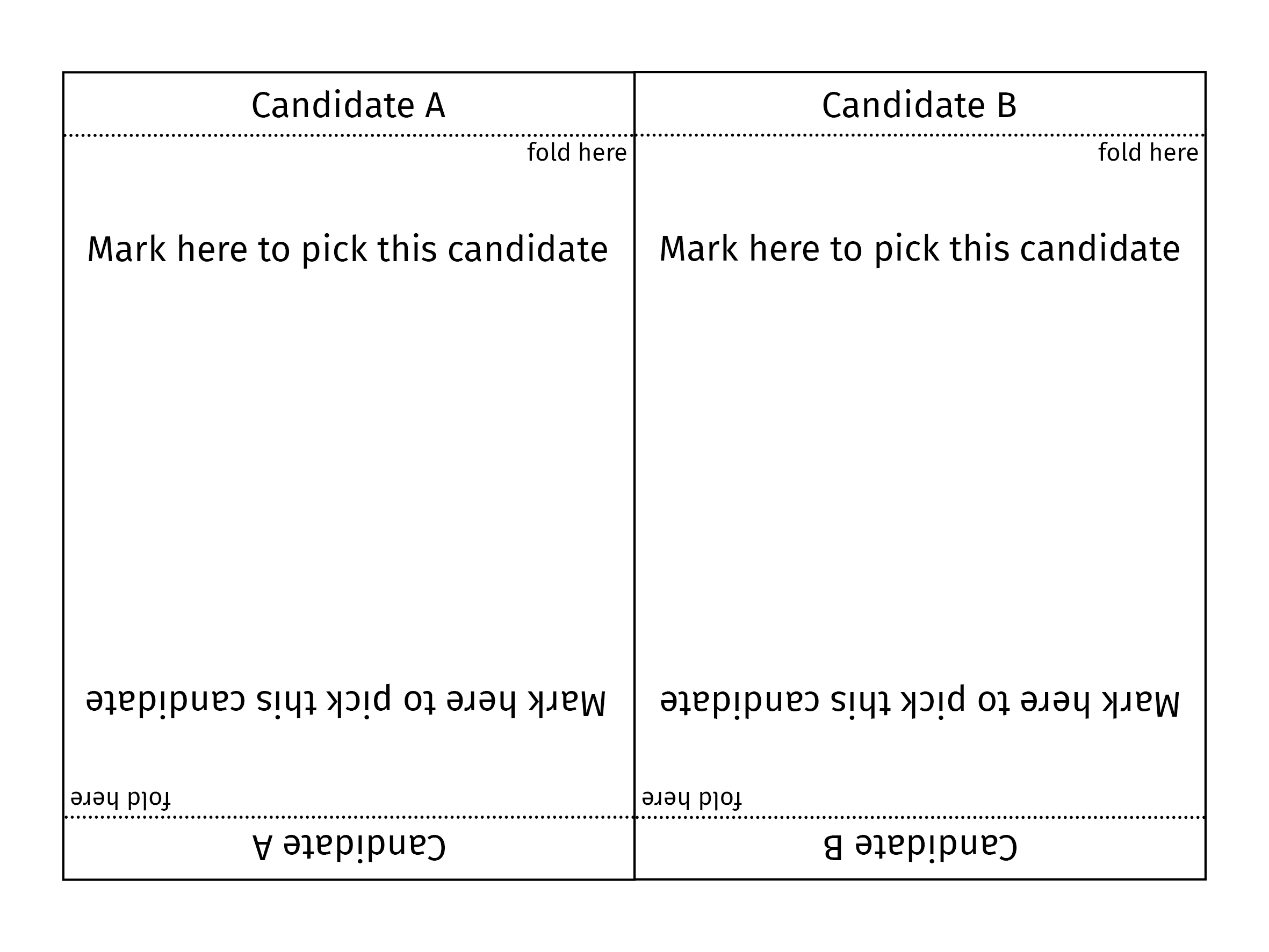}
    \caption[A foldable paper ballot for a choice between two candidates]{A foldable paper ballot enables a voter to mark their ballot with privacy in a crowded boardroom. This ballot is for a binary choice between two candidates.
    The voter makes a mental note of where each label is,  folds the edge of the ballot on top of the label, and puts a mark in the zone corresponding to the candidate of their choice --- all in plain sight of other voters. 
    }
    \label{boardroom:fig:foldedbinary}
\end{figure}

To protect ballot confidentiality during ballot marking, we propose {\it foldable paper ballots}. 
A paper ballot might consist of two labelled columns, where each column corresponds to a particular choice, which is labelled at both the top and bottom of the ballot
(see Figure~\ref{boardroom:fig:foldedbinary}).

To mark a ballot, the voter makes a mental note of the labels for each column 
and folds the top and bottom portions of the ballot down over the labels to hide them. 
They might move or rotate  it in any way that allows them to remember what is under the part of the ballot the will mark. 
When rotating
their ballot, a voter should prevent adversaries from observing the number of manipulations, for example, by rotating the ballot beneath a large cloth. 
The ballot's symmetry adds to the difficulty of an adversary trying to view manipulations the voter makes before marking the outside.

The folding can be temporary or permanent.
To prevent an adversary from unfolding the ballot and glancing at a label, 
part of the paper could be adhesive to make a permanent fold.
To make temporary folding more secure (against quickly unfolding the label), the top and bottom parts could be folded twice. 
A limitation of the foldable ballot 
is that it requires the voter to remember what is inside a part of ballot and might increase the rate of ballot-marking errors.

The foldable ballot can be generalised to support additional ballot choices by using a polygonal paper ballot
or candidate wheel, which is continuous band of paper with candidates on the side.
\vspace*{-6pt}

\vspace*{-6pt}


\vspace*{-12pt}
\subsection{Visual Secrets}

 A voter may use  {\it visual secrets} to enable voters to verify their votes without being able to prove how they voted,
A visual secret is an image or pattern that is easy to recognise but hard to describe.

\begin{figure}
\centering 
   \includegraphics[width=0.40\textwidth]{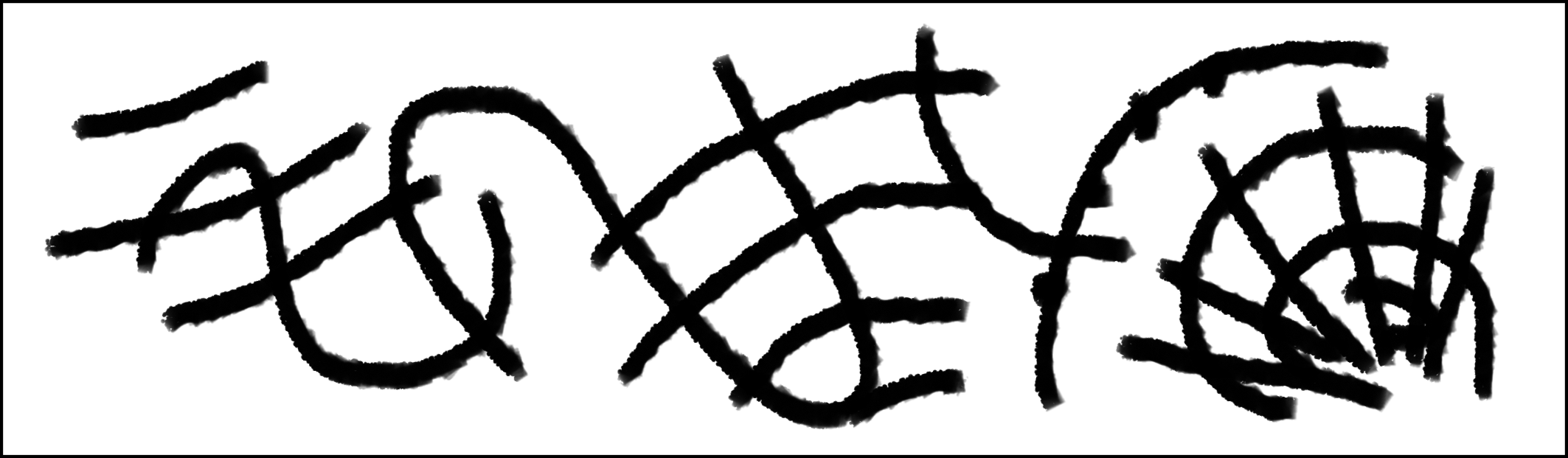}\hfill
   \includegraphics[width=0.40\textwidth]{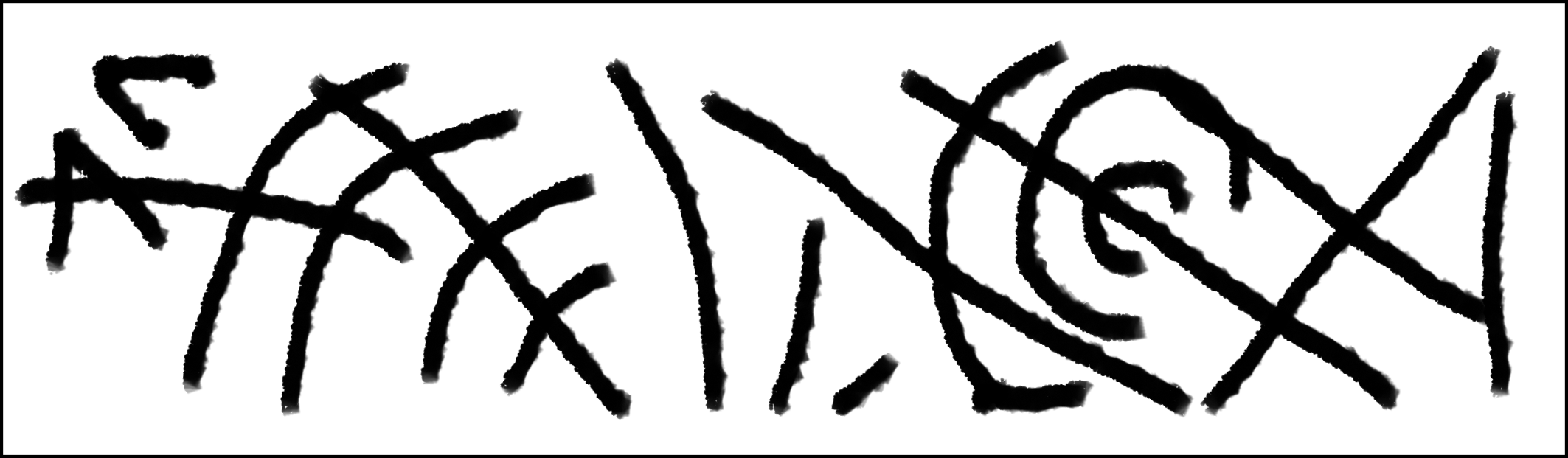}
    \caption[Two examples of visual secrets]{Two examples of visual secrets, which are easy to distinguish and remember but hard to describe orally in a few sentences. These patterns are relatively simple; more complex ones could be used.}
    \label{boardroom:fig:visualsecret}
\end{figure}

The idea is to use a set of images built with similar patterns though visually different.
Each voter could mark their ballot choices with a visual secret.
For example, 30 different images of lions could be taken among a set of 1000, making it hard to describe any image with high precision succinctly. Alternatively, abstract patterns could also be used (see Figure~\ref{boardroom:fig:visualsecret}). 

There are several possible ways of marking a ballot with visual secrets, including with stamps,
peel-off stickers, or images under a scratch-off covering. Self-inking stamps reduce steps and are indelible (see Section~\ref{sec:noveluses}).
Providing each voter with a sheet of peel-off stickers has limitations: the sheet with the remaining stickers reveals which visual secrets were used, and a corrupt or coerced voter could expose the chosen sticker before applying it, The sticker could also potentially be moved or removed.
Ballots with scratch-offs require somewhat complex advance planning.

During tallying, all ballots are revealed (e.g., placed on the center of a large table) and each voter can verify
their ballot, identifying it by their visual secret.  Provided a coerced voter cannot communicate their secret to the 
adversary before the ballots are all revealed, they are safe.  After the ballots are revealed, the coerced voter could
tell the adversary they voted according to any other revealed ballot with the desired ballot choice.

One drawback of visual secrets is that some people might forget the pattern or confuse it for another.
Also, an adversary might view the visual secret being chosen. 
Humans have excellent abilities for visual 
recognition --- better than their abilities to recognise strings --- especially 
with short-term memory~\cite{N.Shepard1967Recognitionmemorywords}.  Another limitation is that it may be possible to
describe the pattern uniquely by describing only a part of it, and doing so might be easier than describing the entire pattern.

\vspace*{-16pt}
\subsection{Parallel Vote Tallying}\label{boardroom:parallelelections}

We explain how to hold multiple parallel tallies for the same election by duplicating ballots, with different guarantors for each ballot box.  Doing so can reduce the trust required in any single authority responsible for the tallying phase.
Voters might suspect that someone could maliciously handle the ballots during this phase.
The main challenge is to ensure that the ballots cast into each ballot box are identical; otherwise,
parallel tallies could facilitate discreditation attacks.  

\begin{figure}   
    \centering 
    \includegraphics[width=0.5\textwidth]{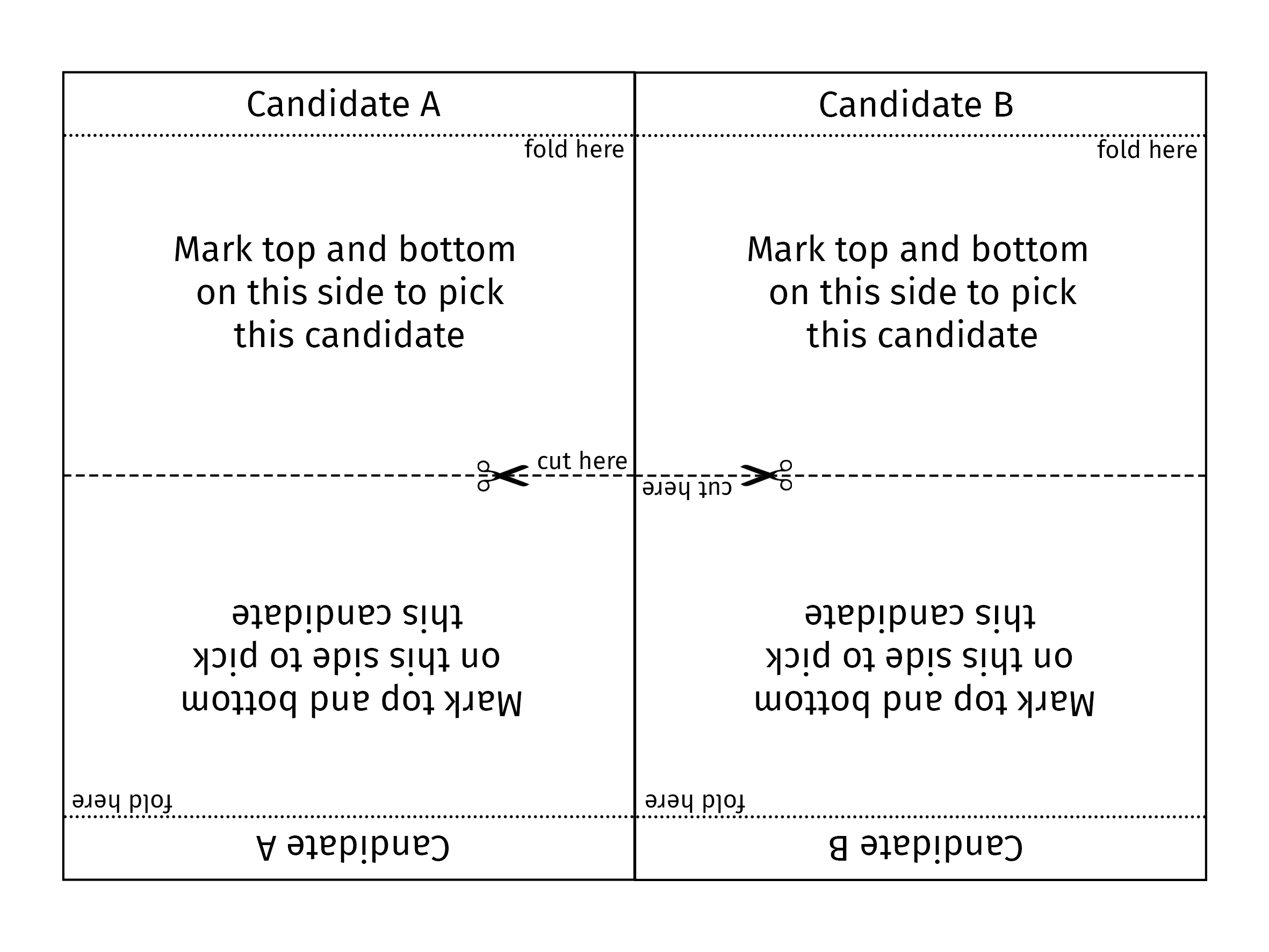}
    \caption[A ballot design for parallel vote tallying]{A ballot design for parallel vote tallying that forces voters to vote for the same candidate in both elections. Each voter makes two marks on the same column, then cuts the ballot in two along the dashed horizontal line, and casts each half in a different ballot box.}
    \label{boardroom:fig:parallelsquare}
\end{figure}

One solution uses a variation of the binary foldable ballot. In this variation, the space where the voter is supposed to make a mark is split into two columns, vertically (see Figure~\ref{boardroom:fig:parallelsquare}). Once the paper is folded, the voter makes two marks on the same column, which can be checked by other people in the room, before cutting the ballot in half and casting each half in a different ballot box. This method can be generalised to more candidates using a variation of the candidate wheel~\cite{Blanchard2019Usabilitylowtech}.

\section{Voting Protocol}
\label{sec:protocol}
\label{boardroom:solutions}

We propose a new paper-based boardroom voting protocol, BVP1, that
offers ballot privacy and voter privacy
when voters are seated around a table.
The protocol combines the building blocks of 
foldable ballots, randomised stamps, random draws, invisible ink, 
and a ballot box on a scale (see Section~\ref{sec:more-blocks}).
We assume there is a single ballot question with
$k$ choices, where $k$ is small enough that a $k$-ary foldable
ballot works (say, $k < 7$).  
We also discuss variations of this protocol.


\subsection{Boardroom Voting Protocol~1 (BVP1)}

We describe {\it Boardroom Voting Protocol 1 (BVP1)} in terms of its setup, ballot marking, casting, counting, and verification steps.
Let $n$ denote the number of voters.  

\bigskip \noindent {\it Setup.} The election authority prepares $n$ or more $k$-ary foldable ballots and an opaque bag of $n$ externally indistinguishable visual-secret stamps, each inked with invisible ink. 
Each stamp imprints a random abstract pattern.
The protocol also requires a ballot box, scale, and one or more opaque black cloths.  The $n$ voters are seated at a table on which there are one or more black cloths. 
To deal with spoiled ballots and stamp malfunctions, the election authority should also prepare some number of extra ballots and stamps.



\bigskip \noindent {\it Ballot Marking and Casting}


\begin{enumerate}

    \item Each voter receives a $k$-ary foldable ballot, where each edge corresponds to a ballot choice.
    
    \item The election authority places $n$ visual-secret stamps in the middle of the table, where the voters can observe that the stamps do not have any externally identifying features.
    
    \item The election authority places the stamps in an opaque bag one-by-one under scrutiny of the voters, after which the bag is slightly shaken and passed around the table. Each voter takes one stamp out of the bag.
    
    \item Each voter visually inspects the pattern on their stamp and remembers it. 
    \item Each voter folds the edges of their ballot and rotates the ballot under the cloth until they are confident that only they know which side corresponds to which candidate.
    
    \item In plain sight, each voter stamps the cell of their choice on their ballot. 
    
    \item One by one, each voter casts their ballot into a ballot box on a scale in a clearly visible place in the room.
  
\end{enumerate}

\bigskip \noindent {\it Counting and Verification}
  
\begin{enumerate}

    \item The election authority shakes the ballot box, takes out the ballots, unfolds them, and places them on a table for all to observe (but not touch).  The election authority sprays revealing ink on the ballots.
    
    \item The election authority counts the number of ballots, checks the vote on each (corresponding to which cell has been stamped), tallies the results, and writes down these numbers for all to see.
    
    \item Each voter verifies the counts and looks for their visual secret.
    
    \item If any voter does not see their visual secret or disputes any count, or has any other concern, they may raise an objection stating their concern.     
    
    
    \item If the number of objections is less than half the margin of victory, the winner is elected. Otherwise, the election is annulled.  A new election can be taken.
    
    
\end{enumerate}

\medskip
Section~\ref{sec:variations} includes a description of an optional procedure for adjudicating claims of missing visual secrets raised in counting and verification Step~4.

\subsection{Variations}
\label{sec:variations}

We discuss four optional variations: voting station, 
rotating ballots under the table, 
parallel ballot collection and tallying,
and protection against discreditation attacks --- which
offer different tradeoffs among complexity, privacy, and outcome integrity.



\paragraph{Voting station.}

Instead of voting at the main table, each voter could vote, one-by-one,
at a dedicated voting station in the room, with observers from different factions. The station might be a table with a stack of ballots, a bag of unused stamps, a ballot box, and an opaque cloth.  
This setup, albeit slower, would provide slightly better privacy and would better accommodate larger sets of voters. 

\paragraph{Rotating ballots under the table.}
Instead of using opaque cloths, voters could rotate their ballots under the table.  This simpler method, however, might make it easier for
malicious voters to exchange ballots in a chain-voting attack (see Section~\ref{sec:attacks}).

\paragraph{Parallel ballot collection and tallying.}

When the environment is highly contentious with high risk of attack,
it may be difficult for the voters to agree on an election authority, and there might be increased risks for discreditation attacks. In such situations, it may be
helpful to conduct the ballot collection and tallying portion of the election in parallel, with each of two factions controlling one ballot box.  

Section~\ref{boardroom:parallelelections} describes a mechanism for ensuring that each voter submits the same ballot choices to each ballot box.  Because BVP1 uses invisible ink, voters would carry out two rounds of stamping: 
first with a common stamp that simply imprints a visible black disk, then second with their unique stamp. Other people in the room can check that each voter stamps two black disks in the same column, and that people only stamp with invisible ink next to a black disk. 





\paragraph{Protection against discreditation attacks using receipt ballots.} 

BVP1 offers only weak voter verification: each voter knows whether or not their ballot was properly collected and counted, but they cannot convince others of this fact.  For example, one or a few voters could falsely
claim that their visual secret is not present or that their ballot
is filled out incorrectly.  BVP1 offers no way to adjudicate such claims, other than to ignore them if their numbers do not affect the election result.  The following variation offers increased protection against discreditation attacks at the cost of diminished ballot privacy.

Using the procedure described in Section~\ref{boardroom:parallelelections} for parallel vote tallying, 
each voter keeps one of the ballots (which we shall call the ``receipt ballot'') on the table in front of them in plain sight.  Observers cannot see the visual secret because it is imprinted with invisible ink.  After the cast ballots are counted, let $j$ be the number of voter raising an objection. If $j$ is less than half the margin of victory, then the objections cannot affect the election outcome.




If $j$ is at least half of the margin of victory, then the following process can be carried out to adjudicate the objections.
The election authority collects all of the receipt ballots in front of voters raising an objection.  After mixing these receipt ballots in an initially empty ballot box, the election authority places them in a central part of the table and sprays them with revealing ink. Then, everyone can compare the revealed receipt ballots with the 
set of cast ballots. An objection is deemed valid if and only if
the associated revealed receipt ballot does not match any other
of the cast ballots.

If the number of validated objections $j\prime$ is at least half the margin of victory, then the election is annulled.

At the end of the election all ballots should be mixed together. They can be saved for election challenges or destroyed to eliminate manipulation before someone challenges it.





\section{Discussion} 
\label{sec:analysis}

We analyse our voting protocol, including
its outcome integrity, ballot privacy, 
usability, and potential vulnerabilities and attacks.


\subsection{Outcome Integrity}
\label{sec:integrity}

The integrity of the election outcome rests on the ballots being
cast as intended, collected as cast, and counted as collected.
All ballots are in plain sight from their distribution until they are shuffled in the ballot box, except for the moment when they are rotated under the cloth (or table). This fact makes it hard for an adversary to modify or replace another voter's ballot.  

Assuming each voter can remember and identify their visual secret, each voter can verify if their ballot has been correctly collected and counted. Although each voter can notice if their ballot has been altered, they cannot prove it (unless using the receipt ballot variation). Because the ballot box sits on a scale, attempts to cast more than one ballot can be detected.  

Threats to outcome integrity include voter mistakes in remembering their visual secret or keeping track of the ballot orientation.  In addition, discreditation attacks might cause the election to be annulled. 


\subsection{Ballot Privacy}
\label{sec:privacy}

The inability of someone in the room to link a voter to a cast ballot depends on several assumptions, including: 
the ability of the voter to hide the orientation of the ballot,
the inability of observers to read the invisible ink,
and the absence of cameras in the room.

In addition, to protect against malicious or coerced users, 
it is important that the voter be unable to: describe their secret in a way that uniquely identifies it,
show their ballot orientation or marks to anyone else,
secretly imprint and exfiltrate their visual secret, 
or make any identifying marks on the ballot.

The receipt ballot variation reduces the anonymity set of those making an objection to the number of people
making objections.


\subsection{Usability}
\label{sec:usablity}

The user experience for an alert sighted voter: 
the voter acquires a ballot, 
folds it, and manipulates  it  under the cloth keeping track
of its orientation. They take a stamp from a bag, look at it to learn the pattern, stamp their ballot in the desired area, and cast the ballot into a ballot box. Throughout the entire voting process, the voter observes activities in the room.

During the counting and verification phase, the voter looks for their
ballot by looking for their visual secret.  After finding their ballot, the
voter verifies that it is marked correctly.
The voter also verifies the tally and the number of ballots counted.

A study is needed to determine how well voters can carry out these tasks.  
Potential difficulties include
keeping track of the orientation of the ballot, 
remembering the visual secret, and being able to notice
possible malicious activities.


\subsection{Potential Vulnerabilities and Attacks}
\label{sec:attacks}

We consider several potential attacks.  Inspired by chain voting~\cite{Saltman2006HistoryPoliticsVoting}, 
an adversary could acquire a stamp, discreetly stamp their own ballot, and exchange their ballot with that of a coerced or bribed voter.  
With the ballots in plain sight, it would be difficult to do so without detection, especially involving many voters.

In an attempt to defeat the variation for imprinting two identical ballots, a malicious voter could feint imprinting one of the invisible ink marks without making an imprint.  To mitigate this threat, part of the stamp (not part of the visual secret) should be inked with visible ink with a simple common mark.

A malicious or coerced voter could make a uniquely identifiable mark on their ballot --- for example, by pricking a pin hole in a certain location, or intentionally smudging the stamp in a certain way.
Similarly, a corrupt election authority could distribute uniquely identifiable ballots with discreetly placed pin holes, marks, or tears. This latter attack can be mitigated by putting the unmarked ballots in a bag and drawing them at random. 

It would be difficult to ensure that there are no miniature hidden cameras in the room or on malicious voters. 
Privacy enhancers partially address this concern, as
it is easier to ensure that the ballot is not in the field of those cameras while under a cloth. 

A malicious or coerced voter could attempt to show the orientation of their ballot to a nearby adversary, or
create a crease that makes the ballot identifiable. 






\subsection{Dealing with Election Failure}
\label{sec:failure}

All election systems are vulnerable to denial-of-service attacks, which can be easy to carry out 
(e.g., bomb or threaten to bomb the voting place).  Similarly, for most election systems, the system cannot prevent attacks on the
election outcome, but at best can detect such attacks.  One advantage of boardroom elections is that, in comparison with large-scale elections, they are relatively easier to re-run if necessary.  Also, in many boardroom contexts, the cost to an adversary of getting caught is especially very high. While re-running an election would be a highly undesirable outcome, this outcome exists as a final option. It then makes sense to have  a secondary highly secure, although possibly less usable, system at hand. 
Voters could then use an easy, fast, and usable protocol that only guarantees detection of fraud (but not necessarily resolution). The existence of a backup solution and deterrence allows voters to benefit from  increased efficiency, while reducing the risk of
election annulment. 

\subsection{Comparison with Paper Ballots}
\label{sec:comparison}

We briefly compare BVP1 with a typical use of simple paper ballots, a protocol we shall call {\it SPB}.
Details can matter greatly in voting protocols, yet SPB rarely is defined by rigorous rules nor carried out in
compliance with such rules. We imagine SPB to work as follows:   The EA distributes simple paper ballots.  Voters mark the
ballots, fold them, and toss them in the center of the conference table.  With the help of the voters, the EA
collects the ballots.  Seated at the table, the EA tallies the ballots and announces the tentative tallies.  
Some voter verifies these tallies.
If there are no objections, the EA declares the results official and maintains the marked paper ballots as evidence.

The main advantages of SPB are simplicity, speed, and low cost.  Assuming everyone can watch everyone else, SPB
also enjoys some outcome integrity.   
The disadvantages of SPB include very poor privacy (it is easy to see how nearby voters vote, and the collected ballots 
are not well mixed).  
Outcome integrity is threatened by
sleight-of-hand by anyone touching the ballots (it would be possible for someone to replace cast ballots without detection).  
As with any system in which voters directly mark ballots, it is virtually impossible to prevent a corrupt 
or coerced voter from intentionally making a unique identifying mark, tear, or fold on the ballot.

It would significantly improve SPB to implement it using a private voting area and ballot box.  For example, there could
be a voting table, with a small privacy screen, in one area of the room where voters could mark their ballots one person at a time.
A large ballot box in plain view of all could rest on the voting table.  At the cost of decreased speed, the private voting area would give voters the option of greatly improved privacy. The ballot box would enhance outcome integrity by not allowing anyone to handle
the ballots before counting.

Because the simpler SPB lacks a voter verification step, it is more susceptible than BVP1 is to attacks that replace ballots.
When SPB is conducted without a private voting area, BVP1 has significantly stronger privacy properties than does SPB.

\subsection{Open Problems}
\label{sec:open}

Open problems include
(1)~Conducting usability tests of the new building blocks and voting protocol.
(2)~Devising additional solutions to boardroom voting
that provide stronger verifiability,
better protection against discredidation attacks, or
greater simplicity.
(3)~Finding solutions that work for voters with visual impairments.

\section{Additional Building Blocks}
\label{sec:more-blocks}

We briefly describe selected additional building blocks, which we use in our protocol or which might be useful in designing other physical protocols.  These building blocks include pre-existing ones (e.g., invisible ink) and 
novel uses of existing mechanisms (e.g., stamps).  

\subsection{Pre-Existing Building Blocks}

Pre-existing building blocks include 
privacy enhancers, locked boxes, transparent ballot boxes, physical commitments,
random-draw methods, cut-and-choose, and invisible ink.


\myheader{Privacy Enhancers.}
Privacy enhancers, such as booths, opaque panels, or pieces of cloth under which voters can manipulate objects, can allow voters to make certain decisions and mark ballots in secret. These devices are especially useful in
boardroom elections, where all voters can observe each other.

\myheader{Locked Boxes.}
Identical small boxes, each with a lock, can be an effective way to encrypt, to ensure
the integrity of items for a certain duration or as they are changing hands, or to make commitments.
Small items, such as ballots, tokens, pens, or stamps, often change hands in boardroom elections, 
creating opportunities for an adversary to steal or alter them. 


\myheader{Transparent Ballot Boxes.}
Transparent ballot boxes can be used to detect if a voter inserts more than one ballot or no ballot.
Inattentive voters could be fooled by two envelopes being cast at the same time. 
Such ballot boxes have the small inconvenience of making it potentially feasible to follow each envelope during the shuffling process, especially when there are few ballots.


\myheader{Physical Commitments.}
Commitments to character strings or images can be made by occluding them with
removable black photographic tape or scratch-off coverings.  Objects can be locked in
boxes.

\myheader{Random-Draw Methods.}
Many secure voting schemes require the generation of random permutations. 
This process can be easily carried out physically.  For example, 
in some board games, players randomly draw items from a bag
(e.g., Scrabble players draw letter tiles).


\myheader{Cut-and-Choose.}
Cut-and-choose is a mainstay auditing procedure~\cite{Chaum1979TR,Chaum1982thesis}. 
It refers to making duplicates of required items, drawing some (either at random or chosen by an auditor), and examining them thoroughly in public to ensure that they have not been maliciously altered. By taking a few items at random, one can ensure with high confidence that, if a large proportion of all items were deficient, this fact would be detected. It can also be used to reveal part of a secret that is split into multiple sections, as does Chaum's protocol for electronic 
cash~\cite{Chaum1983BlindSignaturesUntraceable}. 
The main drawback of such methods is that they add complexity and time, and require more materials (more ballots or tools so that some can be removed and publicly examined). 

\myheader{Invisible Ink.}
Invisible ink can be used to strengthen ballot confidentiality in multiple ways and is used in the Scantegrity voting system~\cite{Chaum2008ScantegrityIIEnd,Chaum2008ScantegrityEndend}.  
We define invisible ink as any ink that is not visible to the human eye without the use of special tools or chemical reactions. 
We also consider time-sensitive invisible ink that automatically becomes visible after a specified period of time, or that disappears after a certain time. Invisible ink limits the risk of onlookers trying to determine what a voter is writing while they are writing. 
It also allows the resulting secret to be kept in plain sight during the rest of the voting protocol, including during  shuffles, reducing opportunities to alter the ballot. 
Invisible ink may have little negative impact on usability, 
but requires more advanced manufacturing and may slightly increase costs.


\subsection{Novel Uses of Existing Mechanisms}
\label{sec:noveluses}

We describe novel uses of three existing mechanisms:
stamps, scales, and polarising filters.



\begin{figure}
    \centering
    \includegraphics[width=0.1\textwidth]{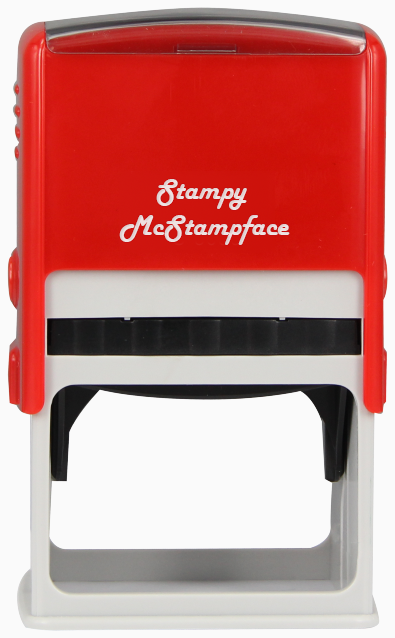}
    \caption{An easily available inexpensive self-inking stamping device.}
    \label{fig:stamp}
\end{figure}

\myheader{Stamps.}
Stamps can be used to imprint special marks on ballots, as needed to implement visual secrets. 
When using stamps to imprint visual secrets, to mitigate the risk of an adversary seeing the pattern,
we recommend two additional precautions.
First, use invisible ink, so that the visual secret is not visible as the voter applies the stamp. 
Second, use a self-inking stamp that rotates when pushed down (see Figure~\ref{fig:stamp}), 
making the pattern visible only when the stamp is pressed. 
With this type of stamp, 
the voter can look at the pattern by pressing it in their hands before their eyes, but 
neighbours cannot see the pattern.
Also, this type of stamp makes showing the pattern to an adversary much more conspicuous. 

Customised stamps can be put in a bag and distributed using a random-draw method.
Care should be taken that the stamps are used only on the ballots and put on the table afterwards, to prevent voters from keeping  proof of how they voted.  

Stamping visual secrets requires custom-made stamps, which are inexpensive.  The stamps
can be re-used a few times, even more so if only a subset of the stamps is taken each time. 
Using visual secrets adds some complexity as it requires two actions by the voter (checking the pattern and stamping the ballot).

\vspace*{-18pt}
\myheader{Scales.}
Putting the ballot box on a mail scale to measure its weight over time can prevent certain attacks 
by detecting if a voter places more than one ballot into the box.  

\myheader{Polarising Filters.}
Another way to reinforce confidentiality is to use polarised light filters, either on ballots, or on a screen used to distribute some common secret. The main advantage of this method is that it makes recording the boardroom with hidden cameras harder, as polarised cameras tend to be bulkier or more expensive~\cite{Prutchi2015DOLPiTwoLow}, and the adversary would need to know the orientation of the polarisation. This method has fewer applications, because it is mostly useful when using a common screen, or small devices that react to polarised light. From a usability standpoint, it requires only polarised glasses, which can easily be found for less than 10\euro, does not significantly increase the time taken to vote, and slightly raises the complexity. 



\section{Conclusion}
\label{sec:conclusion}

This paper introduces the study of secure paper based  boardroom elections
with ballot privacy and voter verifiability using only low-tech cryptography. 
We propose two new building blocks --- foldable ballots and visual secrets --- and protocol BVP1
that uses them. This line of research
is significant because many important elections take place in boardrooms, 
and these elections typically are carried out without ballot privacy or voter verifiability. 
Other modern proposals for boardroom elections depend on complex technology. It is likely that some people and organisations will not be willing to run boardroom elections using complex technology. Despite the imperfections of our initial proposals and the need to test their usability, 
the proposals offer promise that there may be low-tech methods 
that provide greater election integrity and ballot privacy 
than do the simple paper-ballot systems as used in most boardrooms today.

When comparing voting systems, one must consider a variety of factors, including 
efficiency, simplicity, cost, usability, outcome integrity, ballot privacy, receipt freeness, 
coercion resistance, voter verifiability, and resistance to discredidation attacks. 
Whether BVP1 is better overall than the SPB simple paper ballot system (see
Section~\ref{sec:comparison}), depends in part on the context and implementation details.
BVP1 is more complex, depends on the unproven security of visual secrets, 
and is likely more error prone.  SPB has serious privacy limitations
and offers no voter verification other than the ability of voters to watch each other.
The foldable ballot offers a practical way to mark ballots privately 
while seated nearby peering eyes.

For most boardroom voting systems and settings, hidden miniature cameras pose a significant threat to
ballot privacy.  Our foldable ballot prevents hidden cameras from discerning ballot choices by honest voters.

Because boardroom voting happens in a wide range of situations with varying financial, timing, and usability constraints, there are benefits in having a variety of protocols from which to choose. 
The low-tech primitives we introduce, and the BVP1 protocol and its variants, provide a useful first set of solutions that
avoid certain drawbacks of existing E2E systems including their need for complex audits.
We hope that our work will inspire others to discover even better boardroom election solutions, 
for example, achieving stronger verifiability, improved usability, and greater resistance to discreditation attacks.

\bibliographystyle{splncs04}
\bibliography{fullbib}




\end{document}